\documentclass[a4paper,fleqn,usenatbib,useAMS]{mnras}

\usepackage{graphicx}	
\usepackage{amsmath}	
\usepackage{amssymb}	
\usepackage{multicol}        
\usepackage{bm}		
\usepackage{pdflscape}	

\usepackage[T1]{fontenc}
\usepackage{ae,aecompl}
\usepackage{txfonts}
\title[Chemo-dynamical evolution of $r$-process elements]{Early chemo-dynamical evolution of dwarf galaxies deduced from enrichment of $r$-process elements}

\author[Hirai et al.]{Yutaka Hirai,$^{1,2}$\thanks{JSPS Research Fellow}\thanks{E-mail:  yutaka.hirai@nao.ac.jp} 
Yuhri Ishimaru,$^{3}$ 
Takayuki R. Saitoh,$^{4}$ 
Michiko S. Fujii,$^{1}$ 
\newauthor  Jun Hidaka,$^{5,2}$ 
and Toshitaka Kajino$^{2,1,6}$
\\
$^{1}$Department of Astronomy, Graduate School of Science, The University of Tokyo, 7-3-1 Hongo, Bunkyo-ku, Tokyo 113-0033, Japan\\
$^{2}$Division of Theoretical Astronomy, National Astronomical Observatory of Japan, 2-21-1 Osawa, Mitaka, Tokyo 181-8588, Japan\\
$^{3}$Department of Natural Sciences, International Christian University, 3-10-2 Osawa, Mitaka, Tokyo 181-8585, Japan\\
$^{4}$Earth-Life Science Institute, Tokyo Institute of Technology, 2-12-1 Ookayama, Meguro-ku, Tokyo 152-8551, Japan\\
$^{5}$School of Science and Engineering, Meisei University, 2-1-1 Hodokubo, Hino, Tokyo 191-8506, Japan\\
$^{6}$School of Physics and Nuclear Energy Engineering, Beihang University, Beijing 100191, China
}

\date{Accepted 2016 December 19. Received 2016 December 18; in original form 2016 August 14}

\pubyear{2017}

\begin{document}
\label{firstpage}
\pagerange{\pageref{firstpage}--\pageref{lastpage}}
\maketitle
\begin{abstract}
The abundance of elements synthesized by the rapid neutron-capture process ($r$-process elements) of extremely metal-poor (EMP) stars in the Local Group galaxies gives us clues to clarify the early evolutionary history of the Milky Way halo. The Local Group dwarf galaxies would have similarly evolved with building blocks of the Milky Way halo. However, how the chemo-dynamical evolution of the building blocks affects the abundance of $r$-process elements is not yet clear. In this paper, we perform a series of simulations using dwarf galaxy models with various dynamical times and total mass, which determine star-formation histories. We find that galaxies with dynamical times longer than 100 Myr have star formation rates less than $10^{-3} M_{\odot}$ yr$^{-1}$ and slowly enrich metals in their early phase. These galaxies can explain the observed large scatters of $r$-process abundance in EMP stars in the Milky Way halo regardless of their total mass. On the other hand, the first {neutron star merger} appears at a higher metallicity in galaxies with a dynamical time shorter than typical neutron star merger times. {The scatters of $r$-process elements mainly come from inhomogeneity of the metals in the interstellar medium whereas the scatters of $\alpha$-elements are mostly due to the difference in the yield of each supernova.}
Our results demonstrate that the future observations of $r$-process elements in EMP stars will be able to constrain the early chemo-dynamical evolution of the Local Group galaxies.
\end{abstract}



\section{Introduction}
The chemical abundance of extremely metal-poor (EMP) stars ([Fe/H]\footnote{[A/B] = $\log_{10}({N_{\mathrm{A}}}/{N_{\mathrm{B}}})-\log_{10}({N_{\mathrm{A}}}/{N_{\mathrm{B}}})_{\odot}$, where $N_{\mathrm{A}}$ and $N_{\mathrm{B}}$ are number densities of elements A and B, respectively.} $\lesssim -$3) provides us insight into the early evolutionary history of galaxies. In the case of elements synthesized by the rapid neutron capture process ($r$-process elements),
recent high-dispersion spectroscopic observations 
have shown that large star-to-star scatters in the relative abundance ratios of $r$-process elements to iron; [$r$/Fe], exist in EMP stars in the Milky Way (MW) halo 
\citep[e.g.][]{1995AJ....109.2757M, 2000ApJ...544..302B, 2000AJ....120.1841F, 2000ApJ...530..783W,  
2002ApJS..139..219J, 2002A&A...387..560H, 
2004ApJ...607..474H, 2005A&A...439..129B,
2007A&A...476..935F, 2013ApJ...771...67I, 2013ApJ...766L..13A, 
2014AJ....147..136R}. These scatters may indicate that $r$-process elements were produced in specific site(s) before spatial variations of metallicity distribution in the MW halo have been homogenized. According to the hierarchical structure formation scenario, haloes of large galaxies are formed from the accretion of smaller systems \citep[e.g.][]{1978MNRAS.183..341W, 2007ApJ...667..859D, 2008MNRAS.391.1685S, 2016ApJ...818...10G}. The star-to-star scatters would have already formed in the MW sub-haloes.

Although observations of the MW halo stars cannot directly show us how the MW progenitor galaxies evolved, the chemical abundances of the Local Group (LG) dwarf galaxies would provide hints to help us understand the evolutionary histories of the building blocks of the MW. Each dwarf galaxy seems to have a different distribution of $r$-process elements whereas observed abundances of elements lighter than Zn in EMP stars of dwarf galaxies are more or less in reasonable agreement with those of the MW halo \citep{2010ApJ...708..560F, 2010Natur.464...72F}. The enhancements of $r$-process elements of dwarf spheroidal galaxies (dSphs) are slightly lower than those of the MW halo \citep[e.g.][]{2015ARA&A..53..631F}.
On the other hand, ultra-faint dwarf galaxies have significantly depleted $r$-process abundances in general \citep[e.g.][]{2015ARA&A..53..631F} except for the ultra-faint dwarf galaxy Reticulum II, which shows substantial enhancement of Eu \citep{2016AJ....151...82R, 2016Natur.531..610J, 2016ApJ...830...93J}. Although the distribution of [$r$/Fe] seems to differ among dwarf galaxies, the dispersion at the given metallicity {in the sum of different dwarf galaxies} agrees well with that of the MW halo stars.
These observational results imply that the LG dwarf galaxies should have similarly evolved with building blocks of the MW in the early phase.

This $r$-process abundance in the LG galaxies reflect the astrophysical site of $r$-process elements. It is therefore required to clarify the astrophysical site. Binary neutron star mergers (NSMs) are suggested to be one of the most promising astrophysical sites of $r$-process \citep[e.g.][]{1974ApJ...192L.145L, 1976ApJ...210..549L, 1977ApJ...213..225L, 1982ApL....22..143S, 1989Natur.340..126E, 1989ApJ...343..254M}. Nucleosynthesis studies have shown that NSMs can synthesize heavy elements with a mass number over 110 \citep{1999ApJ...525L.121F, 2011ApJ...736L..21R, 2011ApJ...738L..32G, 2012MNRAS.426.1940K, 2013ApJ...773...78B, 2014MNRAS.439..744R, 2014ApJ...789L..39W, 2016ApJ...816...79S}. Recently, the infrared afterglow of the short-duration $\gamma$-ray burst GRB130603B was detected \citep{2013ApJ...774L..23B, 2013Natur.500..547T}. This observation is presumed to be a piece of the evidence that compact binary mergers are the source of $r$-process elements \citep{2013ApJ...775..113T, 2016MNRAS.459...35H}.
 
On the other hand, long NS merger times may make it problematic to explain the star-to-star scatters of $r$-process elements in EMP stars \citep{1990Natur.345..491M, 2004A&A...416..997A}. 
Binary NSs lose their angular momentum by emitting gravitational waves. Observation of binary pulsars reported that all known and possible binary NSs are expected to take more than $\sim$ 100 Myr after the formation of binary systems to merge \citep{2008LRR....11....8L}. Population synthesis calculations have also shown that large fractions of NSMs have merger times longer than 100 Myr \citep{2002ApJ...572..407B, 2006ApJ...648.1110B, 2012ApJ...759...52D, 2014MNRAS.442.2963K}. Several chemical evolution studies, however, pointed out that the observed star-to-star scatters of $r$-process elemental abundance ratios cannot be explained if they assume merger times of 100 Myr as a source of $r$-process elements \citep{1990Natur.345..491M, 2004A&A...416..997A, 2014MNRAS.438.2177M, 2015A&A...577A.139C, 2015MNRAS.452.1970W} because the mean value of [Fe/H] has already reached over $-$2 when the first NSMs occurred in their chemical evolution models \citep[e.g.][]{2000A&A...356..873A}. 

In addition to the problem of long merger times, \citet{2004A&A...416..997A} suggested that the rate of NSMs also make it problematic to explain the observed abundance ratio of $r$-process elements by using their inhomogeneous chemical evolution models. The rate of NSMs in an MW mass galaxy is estimated to be $\sim$ 1 -- 1000 Myr$^{-1}$ from the observed binary pulsars and population synthesis calculations \citep{2010CQGra..27q3001A}. If the rate of NSMs is such low, we need the high yields of $r$-process elements to explain the total abundance of $r$-process elements in the MW. \citet{2004A&A...416..997A} pointed out that extremely higher abundances of $r$-process elements appear even in [Fe/H] $\sim -$1 than that of observations.

Recent studies have proposed solutions to these problems of NSMs in the galactic chemical evolution \citep{2014ApJ...783..132K, 2014MNRAS.438.2177M,  2014A&A...565L...5T, 2015MNRAS.447..140V, 2015A&A...577A.139C, 2015ApJ...804L..35I, 2015ApJ...807..115S, 2015MNRAS.452.1970W, 2015ApJ...811L..10T, 2015ApJ...814...41H, 2016MNRAS.455...17V, 2016ApJ...830...76K, 2016arXiv161002405C}. 
Properly taking account of the mixing of metals would resolve the first problem related to a low NSM rate \citep{2014A&A...565L...5T, 2015ApJ...807..115S, 2015ApJ...814...41H}. Considering the galaxy formation process would resolve the second issue of long merger times. \citet{2015ApJ...804L..35I} have shown that NSMs with a long merger time of 100 Myr can explain the observed ratio of $r$-process elements to iron such as [Eu/Fe] if the MW halo formed via hierarchical mergers of sub-haloes with different star-formation (SF) efficiencies. 

SF efficiencies of progenitor galaxies of the MW are deeply related to the dynamical evolution of galaxies. It is thus required to examine the dependence on the SF efficiencies on dynamical conditions, by using chemo-dynamical simulations of galaxies beyond simple chemical evolution models. \citet{2015ApJ...807..115S} and \citet{2015MNRAS.447..140V} performed hydrodynamic MW-formation simulations beginning from cosmological initial conditions. Their simulations can self-consistently follow the evolutionary history of the MW. However, mass resolutions of their models are not enough to discuss sub-structures of the MW halo.

\citet{2015ApJ...814...41H} have constructed $N$-body/smoothed particle hydrodynamics (SPH) models of dwarf galaxies with higher resolution. Their results suggest that the observed [Eu/Fe] in EMP stars can be accounted for by NSMs with
merger times of $\lesssim 300$ Myr. Low SF efficiency in the early phase of the evolution of dwarf galaxies causes the formation of EMP stars with $r$-process elements which are consistent with observations. They also identified that enrichment of $r$-process elements has a strong connection with the dynamical evolution of galaxies. These simulations were, however, performed for a fixed dark matter halo model producing only one example of the evolutionary history of a dwarf galaxy. 
On the other hand, observational studies suggest diversity of SF histories (SFHs) of the LG dwarf galaxies \citep[e.g.][]{2009ARA&A..47..371T}. We, therefore, perform simulations with the MW progenitor galaxies with different dynamical evolution to examine how the dynamical evolution of galaxies affects the enrichment of $r$-process elements. 

In Section \ref{Method}, we describe our method and models. {In Section \ref{obs}, we compare the properties of our models and the observed dwarf galaxies.} In Section \ref{r-process}, we show the relation between the dynamical evolution of dwarf galaxies and enrichment of $r$-process elements. {Section \ref{MW} discusses the implication to the MW halo formation.} In Section \ref{Summary}, we summarize the main results.
\section{Methods and models}\label{Method}
\subsection{Methods}\label{method}
We perform a series of simulations with an $N$-body/SPH code, \textsc{asura} \citep{2008PASJ...60..667S, 2009PASJ...61..481S}.  \citet{2015ApJ...814...41H} describe details of our code and models. Here, we briefly describe our method and models.

Gravity is calculated using the tree method \citep{1986Natur.324..446B} to reduce the calculation cost of gravitational interaction. For computation of hydrodynamics, we use a standard SPH method \citep[e.g.][]{1977MNRAS.181..375G, 1977AJ.....82.1013L, 1985A&A...149..135M, 1992ARA&A..30..543M}. {To properly treat a contact discontinuity, we need to modify SPH formalism to the density-independent formalism of SPH \citep*{2013MNRAS.428.2840H, 2013ApJ...768...44S, 2015PASJ...67...37Y}. The difference of the treatment of hydrodynamics significantly affects, for example, the entropy profiles of galaxy clusters \citep{2016ApJ...823..144S} and galaxies which experience ram pressure stripping \citep{2014A&A...564A.112N}. However, in the case of isolated galaxy simulations, \citet{2013MNRAS.428.2840H} has shown that time-averaged SFRs are not significantly different with the formalisms of SPH. We thus choose widely used standard SPH method in this study.} We also implement a time-step limiter to correctly treat strong shocks in high-resolution SPH simulations \citep{2009ApJ...697L..99S} and a fully asynchronous split time-integrator for a self-gravitating fluid to reduce the calculation cost \citep{2010PASJ...62..301S}.  We adopt a metallicity-dependent cooling/heating function from 10 K to $10^9$ K generated by \textsc{Cloudy} \citep{1998PASP..110..761F, 2013RMxAA..49..137F}. {Ultra-violet background heating is implemented following \citet{2012ApJ...746..125H}. We also implemented the effect of hydrogen self-shielding following the fitting function of \citet{2013MNRAS.430.2427R}.} 

For star formation, a gas particle is required to satisfy three conditions: (1) $\nabla\cdot\textbf{\textit{v}} <$ 0, (2) $n >$ 100 cm$^{-3}$ , and (3) $T <$ 1000 K, where $\textbf{\textit{v}}$, $n$, and $T$ are the velocity, the number density, and the temperature of the gas particle, respectively. When a gas particle satisfies these conditions, we convert the gas particle to a star particle following the prescription of \citet{1992ApJ...391..502K} to be consistent with Schmidt law \citep{1959ApJ...129..243S}. We set the value of the dimensionless SF efficiency parameter to be $c_{\star}$ = 0.033 following \citet{2008PASJ...60..667S}. \citet{2015ApJ...814...41H} confirmed that the value of $c_{\star}$ does not strongly affect the SF rates (SFRs) of dwarf galaxies as long as we use such high mass resolution and the set of SF threshold.

We adopt the averaged metallicity of 32 nearest neighbour gas particles in the SF region as the metallicity of newly formed star particles {following \citet{2015ApJ...807..115S, 2015ApJ...814...41H} for metal mixing. We assume that metals in SF region are instantaneously mixed following the observation of uniform metallicity distributions in open-star clusters \citep{2007AJ....133..694D, 2007AJ....133.1161D, 2012MNRAS.427..882T, 2013MNRAS.431.1005D, 2010A&A...511A..56P, 2010AJ....140..293B, 2012MNRAS.419.1350R, 2013MNRAS.431.3338R}.  \citet{2016A&A...588A..21R} have shown that the treatment of metal mixing affect the scatters of [$\alpha$/Fe] as a function of [Fe/H]. In Section \ref{obs}, we confirm that the models reproduce the low scatters of $\alpha$-elements observed in EMP stars of the MW halo and dwarf galaxies. \citet{2015ApJ...814...41H} discussed the impact on metal mixing on the abundance of $r$-process elements.}

We treat each star particle as a single stellar population assuming \citet{1955ApJ...121..161S} initial mass function (IMF) of the power law index $-1 + \alpha$ with $\alpha = -1.35$ between a mass range 0.1 $M_{\odot}$ and 100 $M_{\odot}$. Stars more massive than 8 $M_{\odot}$ are assumed to explode as core-collapse supernovae (SNe) following the prescription of \citet*{2008MNRAS.385..161O}. Each SN ejects thermal energy of $10^{51}$ ergs to the surrounding gas particles. Numerical simulations of galaxy formation widely adopt this method \citep[e.g.][]{2008MNRAS.385..161O, 2008PASJ...60..667S, 2012MNRAS.426..140D, 2013MNRAS.432.1989S, 2013MNRAS.436.3031V, 2014MNRAS.445..581H}.  We adopt the nucleosynthesis yields of \citet{2006NuPhA.777..424N} for core-collapse SNe. We also implement the heating of H$_{\mathrm{II}}$ region by young massive stars using P\'EGASE \citep{1997A&A...326..950F}.

Since our aim is to discuss [Eu/Fe] in the MW halo, where the effect of SNe Ia is small, we do not implement effects of Type Ia supernovae (SNe Ia) to reduce the uncertainty related to the modelling of SNe Ia. In this study, we focus on the abundance of $r$-process elements of EMP stars. We thus plot the abundance of $r$-process elements at 1 Gyr from the onset of the major SF. In this phase, the effect of SNe Ia is still small because typical delay times of SNe Ia are $\sim$ 1 Gyr \citep*{2014ARA&A..52..107M}. We adopt the solar system abundances of \citet{2009ARA&A..47..481A}. 

We assume that 1 \% of stars in the mass range of 8 -- 20 $M_{\odot}$ cause NSMs corresponding to the rate of $\sim 5 \times 10^{-5}$ yr$^{-1}$ for an MW sized galaxy. This rate is consistent with the estimate from the observed binary pulsars \citep{2010CQGra..27q3001A}. {We assume each NSM ejects $2\times10^{-5} M_{\odot}$ of Eu following \citet{2015ApJ...804L..35I}. }
 
Merger time distribution is highly uncertain. \citet{2012ApJ...759...52D} have shown that the merger time has a power law distribution between $\sim$10 Myr to 10 Gyr with the index of $-$1 to $-$2. However, the predicted distribution is highly subject to the treatment of common envelope phases which are not well understood. Besides, the number of observed binary pulsars is still too small to constrain the merger time distribution \citep{2008LRR....11....8L}.

The most pessimistic model of \citet{2012ApJ...759...52D} predicts the merger time of 100 Myr. This time-scale is so far the minimum merger time estimated from the observed binary pulsars \citep{2008LRR....11....8L}. 
 EMP stars are thought to {have been} formed in the early phase of galaxy evolution. {These stars which exhibit $r$-process elements in the stellar atmosphere are thus most likely} polluted by NSMs which have the shortest merger times. We, therefore, take the merger time of 100 Myr as a fiducial value {as in the previous studies} \citep{2004A&A...416..997A, 2014ApJ...783..132K, 2014MNRAS.438.2177M,  2014A&A...565L...5T, 2015A&A...577A.139C, 2015ApJ...804L..35I, 2015MNRAS.452.1970W}. \citet{2015ApJ...814...41H} discuss the variations of merger times. We also discuss the effect of varying merger times of NSMs in Section \ref{mergertime}. 
 \subsection{Isolated dwarf galaxy models}\label{models}
\label{dwarf galaxies}
In this study, we assume isolated dwarf galaxy models following \citet{2009A&A...501..189R} and \citet{2012A&A...538A..82R}. We set the total number of particles and the gravitational softening length equal to $2^{19}$ and 7.0 pc, respectively. We take the mass ratio of gas-to-dark matter particle of 0.15 \citep{2014A&A...571A..16P}. Both dark matter and gas particles follow the pseudo-isothermal profile \citep{1991MNRAS.249..523B}, based on the observed dark matter profiles of nearby dwarf galaxies \citep{2011AJ....141..193O, 2015AJ....149..180O},
\begin{equation}\label{pseudo}
\rho (r) =\frac{\rho_{\rm c}}{1+(r/r_{\rm c})^2},
\end{equation}
where $\rho (r)$ is the density of {both dark matter and gas particles}, $\rho_{\rm c}$ is the central density, and $r_{\rm c}$ is the core radius.  The initial values of $\rho_{\rm c}$ ($0.5$ -- 10.0 $\times$ 10$^7$ $M_{\odot}$ kpc$^{-3}$) evolve to stellar central density of $\sim$ 10$^{8-10}$ $M_{\odot}$ kpc$^{-3}$ at 13.8 Gyr from the beginning of the simulation. These values are consistent with central densities of the observed nearby dwarf galaxies \citep[$\sim~10^{6-10}~M_{\odot}$ kpc$^{-3}$,][]{2011AJ....141..193O, 2014ApJ...789...63A, 2015AJ....149..180O, 2015ApJ...808..158B}. 

The total mass of the galaxy ($M_{\rm tot}$) satisfies the following equation,
\begin{equation}\label{totalmass}
M_{\rm tot} = 4\pi\rho_{\rm c}r_{\rm c}^{3}
\left[\frac{r_{\rm max}}{r_{\rm c}}-\arctan\left(
\frac{r_{\rm max}}{r_{\rm c}}\right)\right],
\end{equation}
where $r_{\rm max}$ is the maximum outer radius. 
We fix $r_{\rm max}$ = 7.1 $r_{\rm c}$ \citep{2015ApJ...814...41H} except for model D. Model D has $r_{\rm max}$ = $r_{\rm c}$ = 1 kpc to discuss chemo-dynamical evolution of dwarf galaxies having very high density resulting in a strong inflow at the early evolutionary phase. We summarize parameters of all models in Table \ref{list}. The central density and total mass are the free parameters of our models. We can estimate the dynamical time of the central region of galaxies from the density of the system ($t_{\rm dyn} = \sqrt{3/4\pi G\rho_{\rm c}}$), where $G$ is the gravitational constant. Higher $\rho_{\rm c}$ therefore gives shorter $t_{\rm dyn}$. We can calculate other parameters from Equations (\ref{pseudo}) and (\ref{totalmass}) with the total number of particles of $2^{19}$.

\begin{table*}
\centering
\caption{List of our models. From left to right the columns show model name, total mass, initial central density, mass of one dark matter particle, the mass of one gas particle, maximum radius, core radius, {merger times of NSMs}, and calculated dynamical times of the central density. \label{list}}
\begin{tabular}{lrr|rrrrrrr}
	\hline
Model&
$M_{\mathrm{tot}}$&
$\rho_{\mathrm{c}}$&
$m_{\mathrm{DM}}$&
$m_{\mathrm{gas}}$&
$r_{\mathrm{max}}$&
$r_{\mathrm{c}}$&
{$t_{\rm NSM}$}&
$t_{\rm dyn}$
 \\&(10$^{8}M_{\odot}$)&
 ($10^{7}~M_{\odot}$ kpc$^{-3})$&
 (10$^{3}M_{\odot}$)&(10$^{3}M_{\odot}$)&(kpc)&(kpc)&{(Myr)}&(Myr)\\
 \hline

		A&3.5&0.5&1.1&0.2&7.1&1.0&{100}&114\\
		{A$_{10}$}&{3.5}&{0.5}&{1.1}&{0.2}&{7.1}&{1.0}&{10}&{114}\\
		{A$_{500}$}&{3.5}&{0.5}&{1.1}&{0.2}&{7.1}&{1.0}&{500}&{114}\\
		B&3.5&1.5&1.1&0.2&4.9&0.7&{100}&66\\
		{B$_{10}$}&{3.5}&{1.5}&{1.1}&{0.2}&{4.9}&{0.7}&{10}&{66}\\
		{B$_{500}$}&{3.5}&{1.5}&{1.1}&{0.2}&{4.9}&{0.7}&{500}&{66}\\
		C&3.5&10.0&1.1&0.2&2.6&0.4&{100}&26\\
		{C$_{10}$}&{3.5}&{10.0}&{1.1}&{0.2}&{2.6}&{0.4}&{10}&{26}\\
		{C$_{500}$}&{3.5}&{10.0}&{1.1}&{0.2}&{2.6}&{0.4}&{500}&{26}\\
		D&3.5&10.0&1.1&0.2&1.0&1.0&{100}&26\\
		{D$_{10}$}&{3.5}&{10.0}&{1.1}&{0.2}&{1.0}&{1.0}&{10}&{26}\\
		{D$_{500}$}&{3.5}&{10.0}&{1.1}&{0.2}&{1.0}&{1.0}&{500}&{26}\\		
		E&7.0&0.5&2.3&0.4&8.9&1.3&{100}&114\\	
		F&35.0&0.5&11.3&2.0&15.3&2.2&{100}&114\\
	        \hline

	\end{tabular}

\end{table*}

We have performed the high-resolution simulation to resolve the scale down to $\sim 10$ pc properly. The mass of a star particle in this simulation is $\sim 100~M_{\odot}$. When the mass of a star particle is $\sim$100 $M_{\odot}$, there are $\sim$ 1 SNe in an SSP particle. In this case, the results depend on the way how we sample the IMF \citep{2016A&A...588A..21R}. We randomly sample the IMF following \citet{2008MNRAS.385..161O}. The IMF sampling is significantly important when we discuss the [$\alpha$/Fe] ratio in galaxies because SN produces both $\alpha$-elements and Fe and different mass SNe produce a different ratio of [$\alpha$/Fe]. On the other hand, when we discuss Eu abundance, the effect of IMF sampling is smaller than the $\alpha$-element abundance because Eu yield does not depend on the mass of the progenitors of NSMs \citep{2014ApJ...789L..39W}. We have checked that scatters of [Eu/Fe] do not strongly depend on the resolution of the simulation.

The chemo-dynamical evolution of dwarf galaxies strongly depends on the initial density and total mass \citep{2001MNRAS.327...69C, 2002MNRAS.335..335C, 2008MNRAS.389.1111V, 2009A&A...501..189R, 2012A&A...538A..82R}. The initial density affects the SFHs of dwarf galaxies \citep{2001MNRAS.327...69C}. Isolated dwarf galaxy models with different total masses describe various observational properties \citep{2008MNRAS.389.1111V}. These are therefore two critical parameters to understand in the relation between dynamical evolution and enrichment of $r$-process elements. They affect the final properties of dwarf galaxies such as metallicity and stellar mass \citep{2012A&A...538A..82R}.
 
\begin{figure}
\includegraphics[width=\columnwidth]{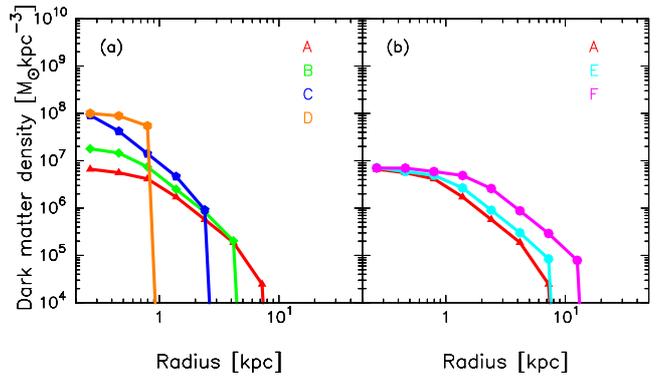}
\caption{Initial density profiles of dark matter particles. (a): Radial dark matter profiles of models with different initial densities. Red, green, blue, and orange curves represent models A, B, C, and D, respectively. (b): The same as those in (a) but for different total masses. Red, cyan, and magenta curves represent models A, E, and F, respectively.\label{RP}}
\end{figure}
 
Fig. \ref{RP} shows the initial radial density profiles of dark matter particles in our models. The initial central density varies by $\sim$ 1 dex among models A, B, C, and D (Fig. \ref{RP}a). On the other hand, models A, E, and F have different values of the total mass, but the central density is the same among them (Fig. \ref{RP}b). In Section \ref{obs}, we show that these models reproduce the observed chemical properties of dwarf galaxies. 

\section{Comparison with observed dwarf galaxies}\label{obs}
{In this section, we compare the predicted and observed properties of dwarf galaxies. Hydrodynamic simulations of galaxies still rely on sub-grid physics such as metal mixing and SN feedback which cannot resolve in simulations of galaxies. It is thus necessary to confirm that our simulations are consistent with fundamental properties of observed galaxies. Here we show that the metallicity distribution function, mass-metallicity relation, and [$\alpha$/Fe] as a function of [Fe/H] predicted in our models are not significantly different from the observed dwarf galaxies.}

\subsection{Metallicity distribution function}
\begin{figure}
\includegraphics[width=\columnwidth]{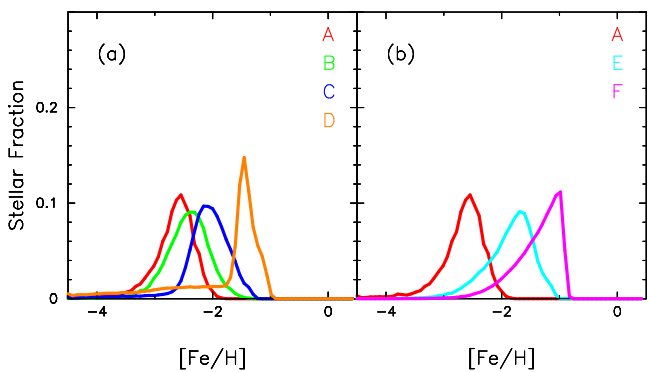}
\caption{{Stellar} metallicity distribution functions at 14 Gyr. (a): Metallicity distribution functions of models with different initial densities. Red, green, blue, and orange curves represent models A, B, C, and D, respectively. (b): The same as those in (a) but for different total masses. Red, cyan, and magenta curves represent models A, E, and F, respectively. \label{MD}}
\end{figure}

Fig. \ref{MD} shows the computed metallicity distribution function (MDF) after 14 Gyr from the beginning of the simulation. We plot the results at 14 Gyr to compare with observational data of the LG dwarf galaxies. Table \ref{final} displays stellar mass, median metallicity, and standard deviations of our models at 14 Gyr. Although we did not try to match the observed dwarf galaxies, standard deviations of MDFs ($\sim$0.35--0.60) lie within that of the observed LG dwarf galaxies \citep[0.32--0.66, ][]{2013ApJ...779..102K}. These MDFs reflect the SFHs of each galaxy. Higher central density models (A, B, C, and D) have steeper slopes of MDFs in low metallicity (Fig. \ref{MD}a). {This result implies that metals are enriched faster in higher central density models in their early phase.} The MDF of Model D shows a narrow peak with low-metallicity tail due to rapid chemical evolution in its early phase. On the other hand, the MDF of model F has a sharp cut-off in the metal-rich side because the SFR does not decrease too much from its peak value. 

\begin{figure}
\includegraphics[width=\columnwidth]{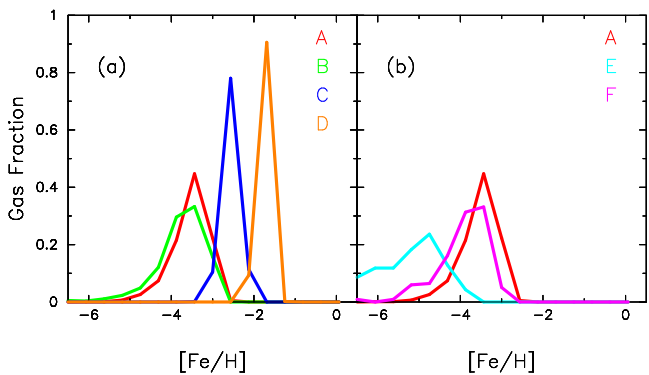}
\caption{{Gas metallicity distribution functions at 0.2 Gyr from the onset of SF. This time roughly corresponds to the time when the first NSM occurs. (a): Gas metallicity distribution functions of models with different initial densities. Red, green, blue, and orange curves represent models A, B, C, and D, respectively. (b): The same as those in (a) but for different total masses. Red, cyan, and magenta curves represent models A, E, and F, respectively.} \label{MDGas}}
\end{figure}

{Fig. \ref{MDGas} shows gas MDFs at 0.2 Gyr from the beginning of the SF. This time roughly corresponds to the time when the first NSM occurs. According to this figure, the gas MDFs of models A, B, E, and F are broader than those in models C, and D. Inhomogeneity of gas phase metallicity significantly affects the scatters of [Eu/Fe] (see Section \ref{density}).}

\begin{table}
	\centering
	\caption{Final properties of our models. From left to right the columns show model name, stellar mass, median metallicity and standard deviation of metallicity distribution function. \label{final}}
	\begin{tabular}{crrr}
		\hline
		Model&
 		$M_{*}$&
 		$\langle$[Fe/H]$\rangle$&$\sigma$\\
		&($10^6~M_{\odot}$)&&\\
		\hline
		A&0.3&$-$2.6&0.35\\
		B&1.2&$-$2.4&0.39\\
		C&3.4&$-$2.0&0.40\\
		D&6.4&$-$1.5&0.60\\
		E&8.0&$-$1.8&0.44\\	
		F&170.0&$-$1.3&0.42\\
		\hline	

	\end{tabular}
\end{table}

\subsection{Mass-metallicity relation}
\begin{figure}
\includegraphics[width=\columnwidth]{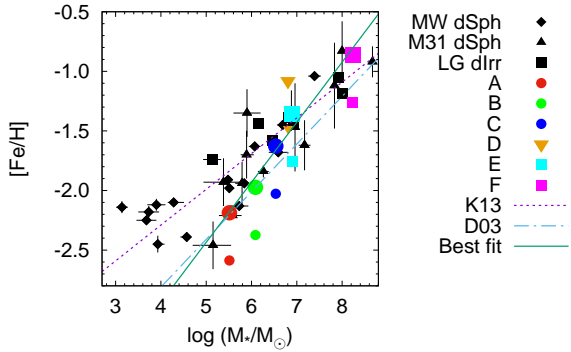}
\caption{Stellar mass-metallicity relation for models and the LG dwarf galaxies. Large, coloured dots stand for 0.4 dex shifted metallicity to account for lack of SNe Ia in models A (red-filled circle), B (green-filled circle), C (blue-filled circle), D (orange-inverted triangle), E (cyan-filled square), and F (magenta-filled square). Small coloured dots stand for the same models, but their metallicity is not corrected. Black diamonds, triangles, and squares with error bars are the observed value of the MW dSphs, M31 dSphs, and the LG dwarf irregulars (dIrrs), respectively \citep[K13]{2013ApJ...779..102K}. The sky-blue dot-dashed line represents the least square fitting of the samples of \citet[D03]{2003MNRAS.344.1131D}: [Fe/H] $\propto M_{*}^{0.40}$. The purple dashed line represents the least-square fitting of K13: $\langle[\rm{Fe}/H]\rangle = (-1.69\pm0.04) + (0.30\pm0.02)\log\left(M_{*}/10^6 M_{\odot}\right)$. The green line represents the least-square fitting of models except for model D: $\langle[\rm{Fe}/H]\rangle = (-1.93\pm0.05) + (0.50\pm0.04)\log\left(M_{*}/10^6 M_{\odot}\right)$.\label{MZR}}
\end{figure}

In Fig. \ref{MZR}, we show the comparison of the predicted mass-metallicity relation to that of the observed dwarf galaxies. The median metallicity in our models increases as the stellar mass increases. This feature is consistent with the observation \citep{2013ApJ...779..102K}. Both metallicity and stellar mass increase as the density (models A, B, C, D) or total mass (models A, E, F) increases. 
More massive or higher-density galaxies have deeper gravitational potential well than less massive or lower-density galaxies. In such galaxies, gas and metals are barely removed from the galaxies \citep[e.g.][]{1986ApJ...303...39D}. In this sense, total mass and density have similar effects on the final stellar mass and metallicity. 

Slightly lower metallicity by $\sim$ 0.4 dex in our models than that in the observations is due to the lack of SNe Ia. According to the observed [$\alpha$/Fe] as a function of [Fe/H] in the MW halo stars, SNe II and SNe Ia produce $\sim$ 35--40 \% and $\sim$ 60--65 \% of solar Fe \citep{2000A&A...359..191G, 2008A&A...489..525P}. Note that the increase of metallicity due to SNe Ia in different galaxies is not a constant value. It is affected by SFHs of each galaxy. However, since the aim of this paper is to discuss $r$-process elements produced in the early phase of galaxy evolution where the contribution of SNe Ia is negligible, discussion of the effect of SNe Ia is beyond the scope of this paper. We thus simply shift the metallicity with 0.4 dex in each galaxy to compare with the observation as shown in large symbols in Fig. \ref{MZR}.

Metallicity in model D is higher than that of the observed mass-metallicity relation. Metals in model D are efficiently retained inside the galaxy because strong inflow caused by a very compact density profile (Fig. \ref{RP}) in the early phase prevents metal-outflow. This model is too extreme as an initial condition for the dwarf galaxies, and indeed the resulting stellar mass and metallicity deviate from the observed relation. From the viewpoint of the mass-metallicity relation, we can exclude model D from the models that represent the observed LG dwarf galaxies.

The slope of the mass-metallicity relation given by models in Fig. \ref{MZR} is slightly steeper than that of the observation. This result might be related to the lack of tidal disruption effect, which we did not implement in this study, and insufficient feedback. The stellar mass of small galaxies can be easily reduced by tidal disruptions \citep[e.g.][]{2014A&A...564A.112N}. Stellar feedback also affects the slope of the mass-metallicity relation. In our simulations, we only assume thermal feedback from SNe. \citet*{2014PASJ...66...70O} point out that the radiation-pressure feedback from young stars suppresses SF more efficiently in galaxies with higher metallicity and produces a shallower slope of the mass-metallicity relation. Several semi-analytic models, as well as hydrodynamic simulations, report the steeper slope \citep[e.g.][]{2013MNRAS.434.2645D, 2014ApJ...795..123L, 2014MNRAS.438.1985T, 2015ARA&A..53...51S}. Isolated dwarf galaxy models adopted in this study might, however, oversimplify the evolutionary history of dwarf galaxies. We will study the evolutionary history of the LG galaxies in the context of standard $\Lambda$ CDM cosmology in a forthcoming paper.
\subsection{$\alpha$-element abundance}
\begin{figure*}
\includegraphics[width=16cm]{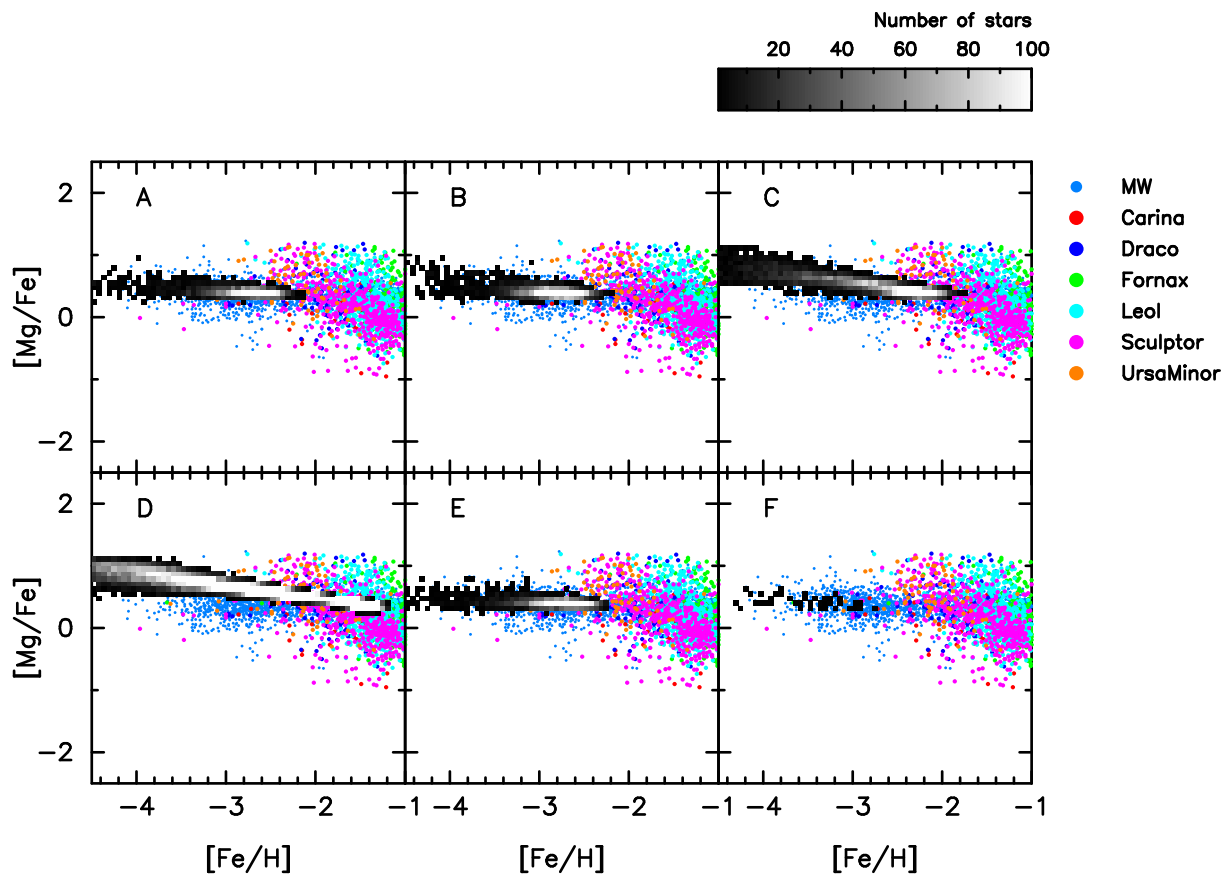}
\caption{[Mg/Fe] as a function of [Fe/H] {for stars} predicted by models after 1 Gyr evolution, i.e. before the onset of SNe Ia. The time of the onset of SF from the start of the simulation in models A, B, C, D, E, and F is 0.21 Gyr, 0.06 Gyr, 0.02 Gyr, 0.02 Gyr, 0.27 Gyr, and 0.10 Gyr, respectively. We plot models A, B, C and D, E, F from left to right in top and bottom panels, respectively. Gray scales are the number of stars predicted in our models, between 1 (black) to 100 (white). The small dots are the observed value of the MW halo stars. Red, blue, green, cyan, magenta and orange dots are the observed value of Carina, Draco, Fornax, Leo I, Sculptor, and Ursa Minor dwarf galaxies, respectively. We compile all data in SAGA database \citep{2008PASJ...60.1159S, 2011MNRAS.412..843S, 2014MmSAI..85..600S, 2013MNRAS.436.1362Y}. \label{MgFe}}
\end{figure*}

{Fig. \ref{MgFe} shows [Mg/Fe] as a function of [Fe/H] for stars before the onset of SNe Ia (after 1 Gyr from the onset of the star formation). The scatters in the observed [Mg/Fe] in EMP stars are less than 0.2 dex.  According to Fig. \ref{MgFe}, all models reproduce the low scatters of [Mg/Fe]. Besides, the metallicity of all models except for model D is [Fe/H] $\lesssim -$2 at 1 Gyr. These results suggest that SNe Ia start to contribute at around [Fe/H] $\sim -$2 in these models, i.e. [Mg/Fe] begins to decrease at around this metallicity which is consistent with the observed MW dSphs. On the other hand, the metallicity of model D is already too high to explain [Mg/Fe] in the MW dSphs. Decreasing trend of [Mg/Fe] in models C and D reflects the adopted yield that more massive stars produce higher [Mg/Fe] ratio \citep{2006NuPhA.777..424N}.} 

\begin{figure}
\includegraphics[width=\columnwidth]{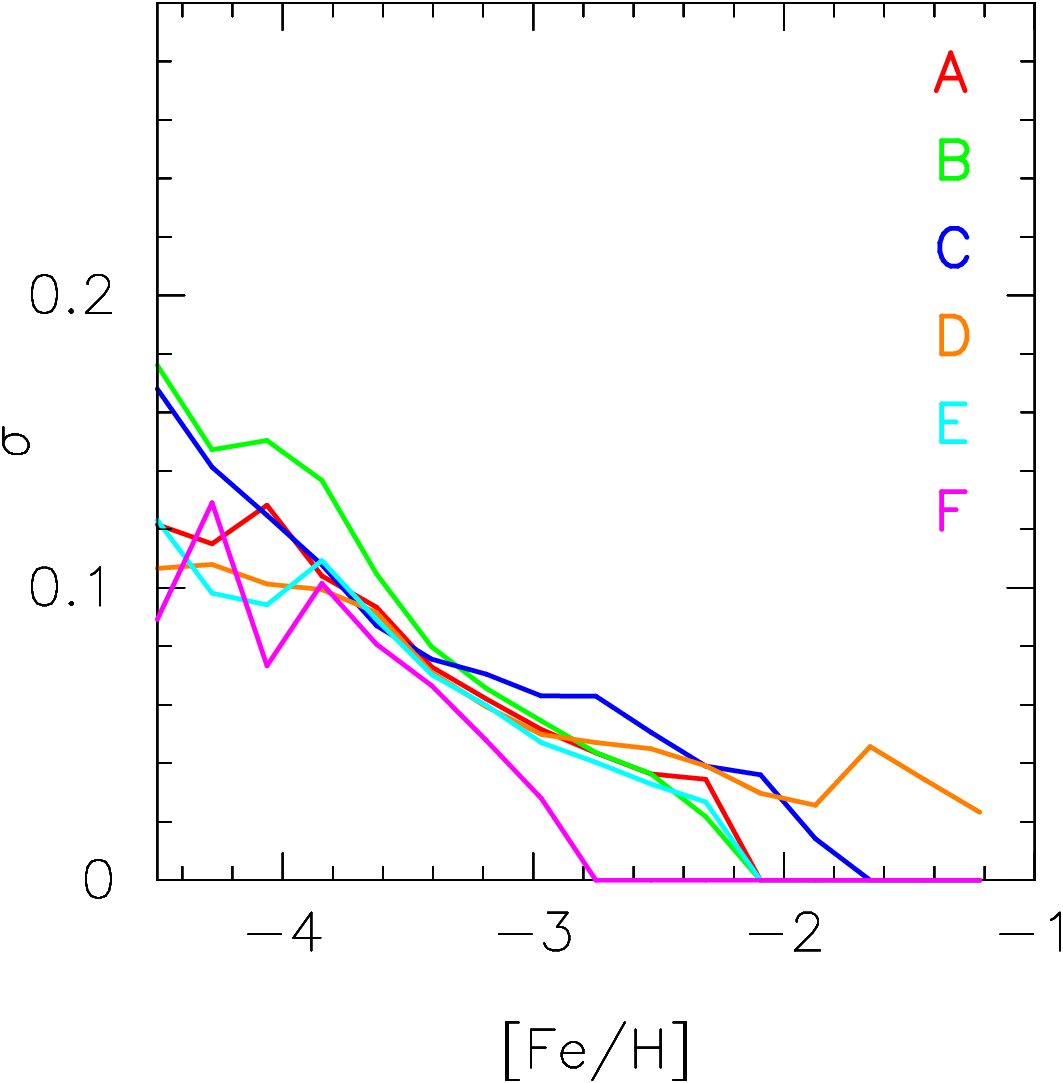}
\caption{Standard deviation of stellar [Mg/Fe] ($\sigma$) as a function of [Fe/H] after 1 Gyr from the onset of SF. Red, green, blue, orange, cyan, and magenta curves represent models A, B, C, D, E, and F, respectively. \label{SigmaMg}}
\end{figure}

\begin{figure}
\includegraphics[width=\columnwidth]{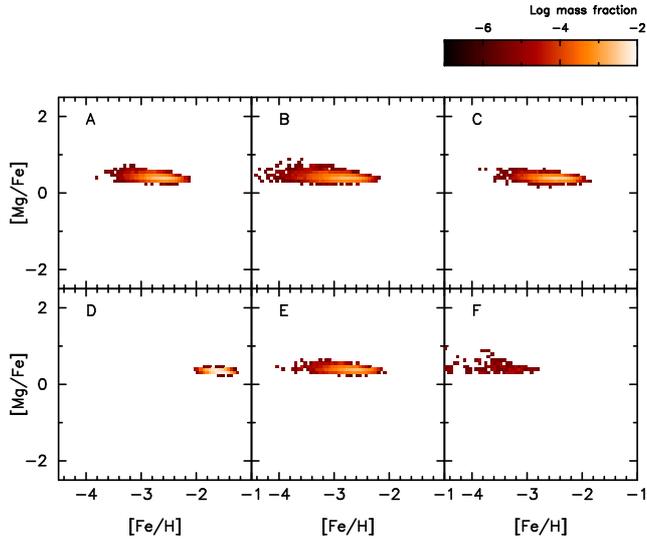}
\caption{{[Mg/Fe] as a function of [Fe/H] for gas computed by models at 1 Gyr from the onset of SF. We plot models A, B, C and D, E, F from left to right in top and bottom panels, respectively. Contours are the mass fraction of gas.} \label{MgFeGas}}
\end{figure}

Fig. \ref{SigmaMg} shows the standard deviation of [Mg/Fe] as a function of [Fe/H]. We can confirm that scatters of all models have lower than 0.2 dex. Standard deviations of all models decrease as the metallicity increases. This result reflects that the spatial distribution of metallicity becomes homogeneous in higher metallicity. 
{Fig. \ref{MgFeGas} shows [Mg/Fe] as a function of [Fe/H] for gases at 1 Gyr. According to this figure, the scatters of [Mg/Fe] are also small in the interstellar medium (ISM) of our models.} According to these results, we confirm that all models have low [Mg/Fe] scatters in EMP stars. This result ensures that the metal mixing scheme adopted here is not too unnatural. 

{The small scatters of [Mg/Fe] mainly come from the difference in the yields of each SN rather than the spatial inhomogeneity of the abundance of Mg and Fe. Since SNe synthesize both Mg and Fe, the spatial variations of the abundance of Mg and Fe are related. When the abundance of Fe is high in a region, Mg is also high in many cases. Therefore, they offset each other due to the above relation making the resultant scatters of [Mg/Fe] very small. }

\mbox{}\\

\section{Enrichment of $\lowercase{r}$-process elements}\label{r-process}
\subsection{Effects of initial density of haloes}\label{density}
\begin{figure}
\includegraphics[width=\columnwidth]{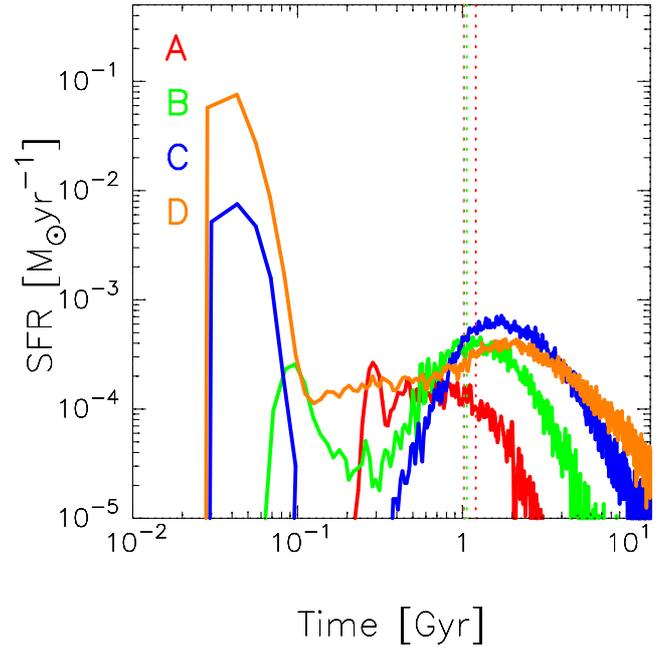}
\caption{SFRs as a function of time. Red, green, blue, and orange curves show models A, B, C, and D, respectively. Vertical dotted lines show the time 1 Gyr after the formation of first stars in each model. These lines correspond to the time when Fig. \ref{EuFe1} is plotted. Vertical dotted lines of model C (blue) and D (orange) are overlapped.\label{SFR}}
\end{figure}

The difference in dynamical evolution directly affects the time variations of SFRs which strongly influence the enrichment of $r$-process elements. Here we show the effects of initial density of haloes (models A, B, C, and D). Fig. \ref{SFR} shows the SFHs in models A, B, C, and D. The gases collapse during the initial dynamical time, $t_{\rm dyn}$, and then the first SF occurs. Dynamical times in models A and B are much longer than the lifetimes of SN progenitors ($\sim$ 10 Myr). This fact means that SN feedback from the first generation stars heats the ISM to prevent the subsequent collapse of gases and the formation of the next generation stars. The SFR in the early epoch is therefore as low as $\lesssim 10^{-3}~M_{\odot}$yr$^{-1}$ in models A and B (Fig. \ref{SFR}), which is consistent with the average observed SFR of the LG dwarf galaxies estimated from colour-magnitude diagrams \citep[e.g.][]{2012A&A...539A.103D, 2012A&A...544A..73D, 2014ApJ...789..147W}.

{On the other hand, dynamical times in models C and D are comparable to the time-scale of the first SN feedback. In these models, SFRs are not suppressed by the SN feedback due to short dynamical times. The SFR in the early epoch thus significantly rises as shown in Fig. \ref{SFR}. When the contribution of SN feedback increases as the time passes, the SFR is suppressed in these models.}

\begin{figure}
\includegraphics[width=\columnwidth]{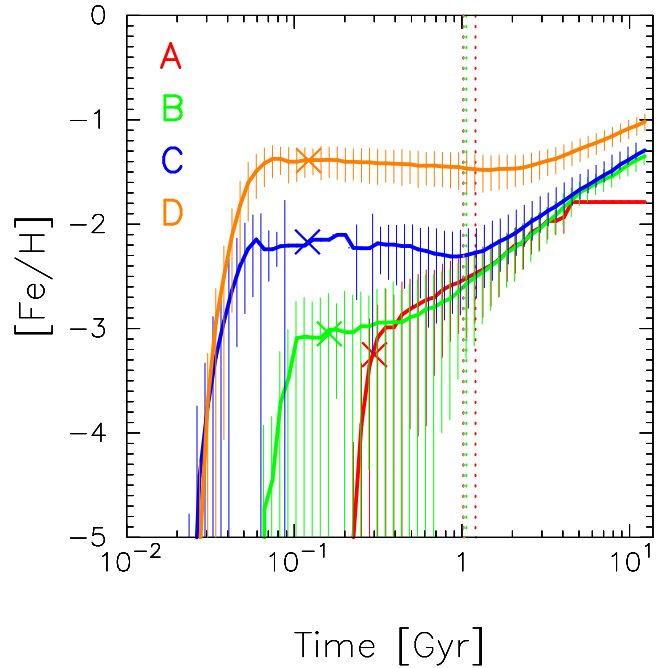}
\caption{The mean stellar [Fe/H] as a function of time (solid curves). Red, green, blue, and orange curves show models A, B, C, and D, respectively. Vertical lines are the dispersion of 2$\sigma$. We plot these lines when the number of stars is more than two. {Around 0.1 Gyr, there is only one star in model C in each bin.} Crosses indicate the value of [Fe/H]$_0$. Vertical dotted lines show the time 1 Gyr after the formation of first stars in each model. These lines correspond to the time when Fig. \ref{EuFe1} is plotted. Vertical dotted lines of model C (blue) and D (orange) are overlapped.\label{AMR}}
\end{figure}

In Fig. \ref{AMR}, we show the mean stellar [Fe/H] in each model as a function of time. The cross symbol indicates the mean value of [Fe/H] at 100 Myr after the first SF takes place in each model. This metallicity corresponds to the metallicity at which the first NSM occurred. Hereafter, we denote this value of metallicity as [Fe/H]$_0$.  As shown in Fig. \ref{AMR}, the value of [Fe/H]$_0$ increases from lower to higher density models, i.e. [Fe/H]$_0 = -3.2, -3.0, -2.2$ and $-$1.4 for models A, B, C, and D, respectively. These metallicities roughly correspond to those at which the large dispersion in [Eu/Fe] appears. The value of [Fe/H]$_0$ depends on the rate of increase in metallicity before the first NSM occurs. Higher SFRs of models C and D increase metallicity before the first NSM occurs resulting in a higher value of [Fe/H]$_0$ as shown in Fig. \ref{AMR}. Metallicities in models C and D become almost constant after the time for [Fe/H]$_0$ due to suppressed SFRs.

\begin{figure*}
\includegraphics[width=18 cm]{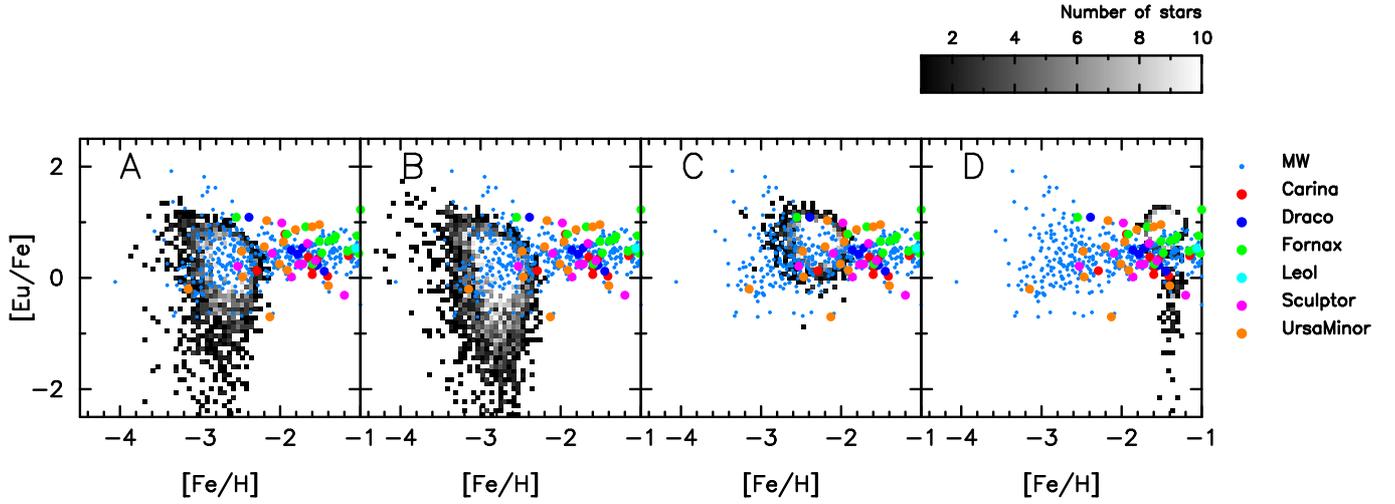}
\caption{[Eu/Fe] as a function of [Fe/H] {for stars} predicted by models with different central density after 1 Gyr from the onset of SF. We plot models A to D from left to right. Gray scales are the number of stars predicted in our models, between 1 (black) to 10 (white). The small dots are the observed value of the MW halo stars \citep[e.g.][]{2014AJ....147..136R}. {Red, blue, green, cyan, magenta, and orange dots} are the observed value of 
Carina \citep{2003AJ....125..684S, 2012ApJ...751..102V, 2012A&A...538A.100L}, 
Draco \citep{2001ApJ...548..592S, 2009ApJ...701.1053C},  
Fornax \citep{2012A&A...538A.100L}, 
Leo I \citep{2003AJ....125..684S},
Sculptor \citep{2003AJ....125..684S, 2005AJ....129.1428G}, and 
Ursa Minor \citep{2001ApJ...548..592S, 2004PASJ...56.1041S, 2010ApJ...719..931C} dwarf galaxies, respectively. All data are compiled in SAGA database \citep{2008PASJ...60.1159S, 2011MNRAS.412..843S, 2014MmSAI..85..600S, 2013MNRAS.436.1362Y}. \label{EuFe1}}
\end{figure*}

Fig. \ref{EuFe1} shows [Eu/Fe] of stars formed in our simulations as functions of [Fe/H] at 1 Gyr from the beginning of the SF, i.e. before the onset of SNe Ia. From models A to C, as the central density increases, the average [Fe/H] at which the first NSM occurs shifts from lower to higher metallicity and the dispersion of [Eu/Fe] decreases. Stars enriched by $r$-process elements appear at around [Fe/H] $\sim~-$3 in models A and B. Few EMP stars are however formed in model C. This means that star-to-star scatters of the $r$-process elements at [Fe/H] $\sim~-$3 seen in the MW halo do not come from dwarf galaxies like those in model C. These galaxies contribute to the stars in higher metallicity stars in the MW halo. Model D shows the scatters of [Eu/Fe] in [Fe/H] $>-2$, which contradict the observations. This result indicates that the assumption of an extremely flat initial dark matter profile (Fig. \ref{RP}) is unrealistic. We can, therefore, exclude model D from the MW progenitor galaxies at least in terms of Eu abundance.

\begin{figure}
\includegraphics[width=\columnwidth]{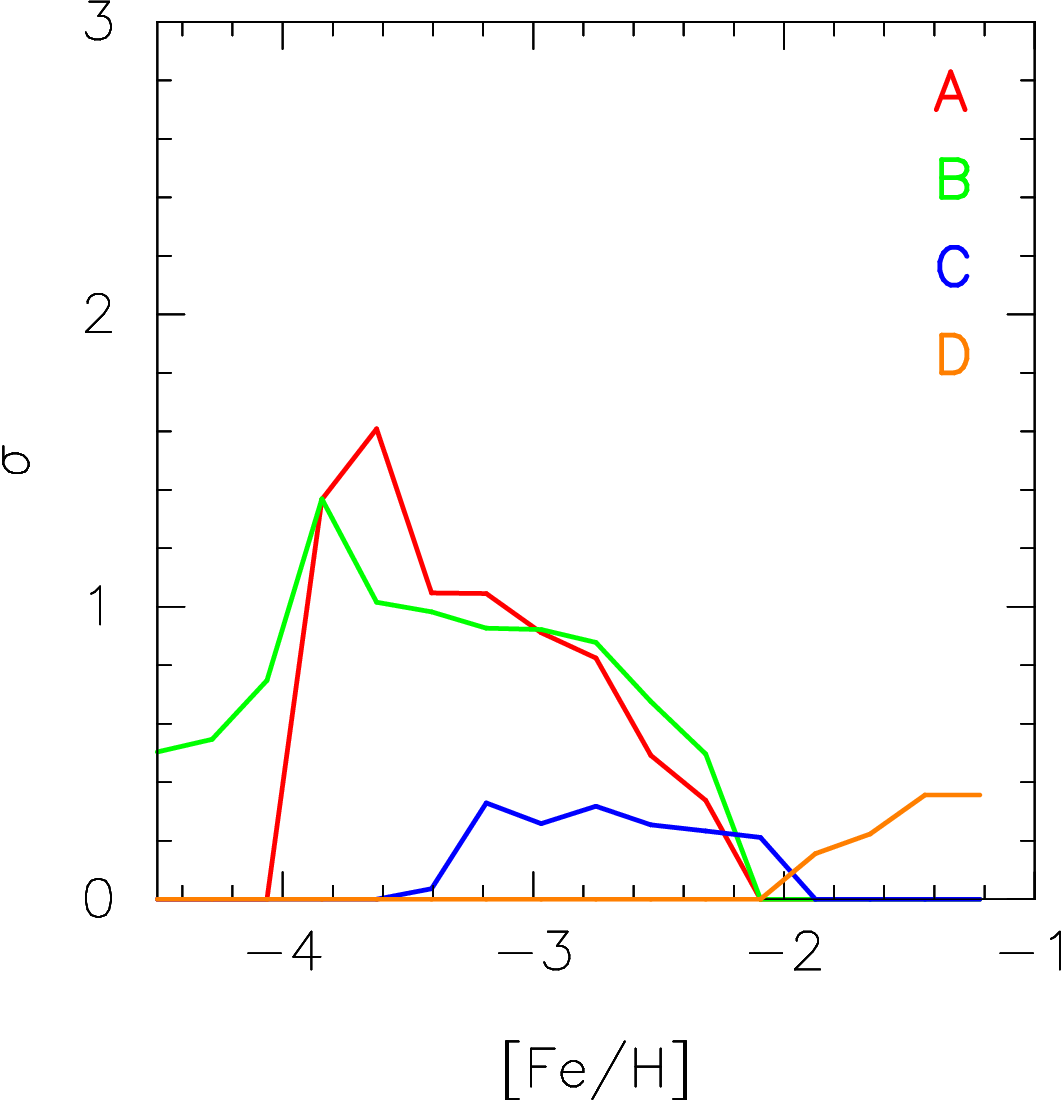}
\caption{Standard deviation of [Eu/Fe] ($\sigma$) as a function of [Fe/H] after 1 Gyr from the onset of SF. Red, green, blue, and orange curves represent models A, B, C, and D, respectively. \label{SigmaEuABCD}}
\end{figure}

{The star-to-star scatters of [Eu/Fe] can come from inhomogeneous spatial distribution of metals in ISM of dwarf galaxies \citep{2015ApJ...814...41H}. Fig. \ref{SigmaEuABCD} shows the standard deviation of [Eu/Fe] ($\sigma$) as a function of [Fe/H]. Models A and B in Fig. \ref{SigmaEuABCD} lead to larger dispersion ($\sigma > 1$) than model C ($\sigma < 0.4$) in [Fe/H] $\lesssim -$2.5. {As shown in Fig. \ref{MDGas}, models A and B have broader gas MDFs than those of models C and D.} These results suggest that the spatial distribution of metals is still inhomogeneous in models A and B when NSMs have first occurred whereas the ISM must have been already homogenized in model C.}

{To quantify the inhomogeneity of metallicity in galaxies, \citet{2000A&A...356..873A} introduced a metal pollution factor ($f_{\rm poll}$). We define it with $M_{\rm poll}/M_{\rm tot}$, where $M_{\rm poll}$ is the gas mass polluted by SNe and $M_{\rm tot}$ is the total gas mass in the galaxy. According to \citet{2000A&A...356..873A}, $M_{\rm poll}$ is equal to $N_{\rm SN}M_{\rm sw}$, where $N_{\rm SN}$ is the number of SNe and $M_{\rm sw}$ is the gas mass swept up by an SN. In our case, we need to add the term of metal mixing in an SF region to $f_{\rm poll}$ defined in \citet{2000A&A...356..873A}. The pollution factor in this study is equal to  $N_{\rm SN}M_{\rm sw}$ + $N_{\star}M_{\rm mix}$, where $M_{\rm sw}$ is the gas mass within SPH kernel when we distribute metals, $N_{\star}$ is the number of stars, and $M_{\rm mix}$ is the gas mass used for metal mixing in an SF region. If $f_{\rm poll} \gtrsim 1$, the whole galaxy is polluted by SNe, i.e. the spatial distribution of metallicity in galaxies is homogeneous. On the other hand, if $f_{\rm poll} \lesssim 1$, the spatial distribution of metallicity is still inhomogeneous, i.e. scatters appear in [Eu/Fe] as a function of [Fe/H]. When we calculate $f_{\rm poll}$ in our models, the values of $f_{\rm poll}$ at 0.2 Gyr from the beginning of the SF are 0.71 (model A), 0.26 (model B), 2.5 (model C), and 41.1 (model D). These results mean that model C and D which have SFRs of $\gtrsim~10^{-2}~M_{\odot}$yr$^{-1}$ in the early phase is already homogeneous in metallicity when the first NSMs occurred. }

The observed dwarf galaxies such as Carina, Draco, Fornax, and Sculptor contain $r$-process elements in EMP stars, although the number of observed EMP stars is insufficient to allow us to discuss the star-to-star scatters of $r$-process elements \citep[e.g.][]{2015ARA&A..53..631F}. The early SFRs of these galaxies inferred from the colour-magnitude diagrams are $\sim~10^{-3}~M_{\odot}$yr$^{-1}$ \citep{2012A&A...539A.103D, 2012A&A...544A..73D}. These values are consistent with models A and B. This result implies that the initial dynamical time of these dSphs might have been of $\sim$ 100 Myr.

As discussed in Section \ref{density}, if the SFR is high such as model C, only high-metallicity stars are enriched by $r$-process. The Sagittarius dwarf galaxy is a candidate for galaxies like model C because it has $r$-process elements only in [Fe/H] $> -1$ if we can sample enough number of stars which have been a member of the Sagittarius dwarf galaxy since it was formed. Most stars in the Sagittarius dwarf galaxy are older than 5 Gyr \citep*{2015MNRAS.451.3489D}. 
The Sagittarius dwarf galaxy might have formed with a high initial density and formed stars with a high SFR. 
However, the number of observed stars is not sufficient to permit us to reach a conclusion. We need statistics of the stars at [Fe/H] $<-1$ to confirm that $r$-process elements are absent in this galaxy.

According to the above discussion, we find that the initial density of haloes of galaxies significantly affects the early SF and the enrichment of $r$-process elements. Although we should confirm this in cosmological simulations of the formation of dwarf galaxies, we expect that their evolutions are similar to our model results once they become a self-gravitating system.
\subsection{Effects of total mass of haloes}\label{mass}
According to the hierarchical merging paradigm, the MW halo is formed via accretions of sub-haloes with different masses. The total mass of each galaxy, as well as the density, is also an important parameter which affects the final metallicity and stellar mass of galaxies  \citep{2012A&A...538A..82R}. Thus, it is necessary to examine the enrichment of $r$-process elements in dwarf galaxies with different masses. 

\begin{figure}
\includegraphics[width=\columnwidth]{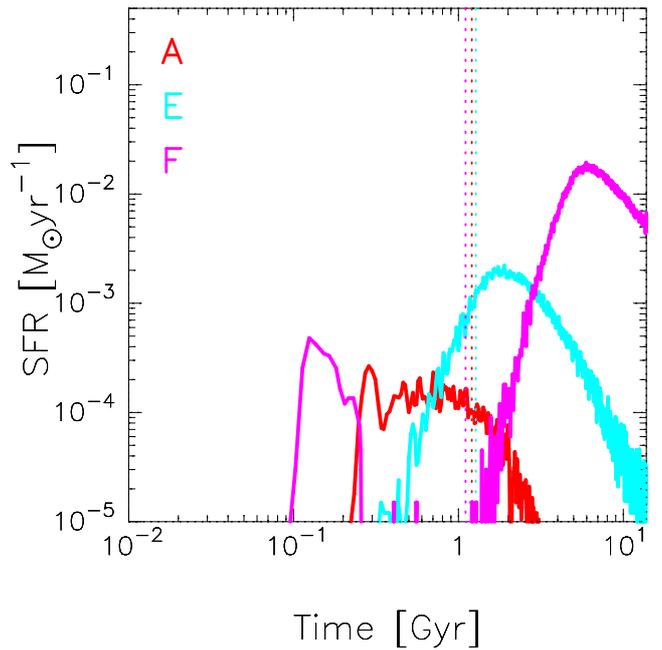}
\caption{SFRs as a function of time. Red, cyan, and magenta curves show models A, E, and F, respectively. Vertical dotted lines show the time 1 Gyr after the formation of first stars in each model.  \label{SFRmass}}
\end{figure}

Fig. \ref{SFRmass} shows time variation of SFRs in models A, E, and F. SN feedback suppresses SFRs in all these models in the early phase. Their dynamical times are of order $\sim$ 100 Myr, which is longer than the lifetime of massive stars. In these models, SN feedback can heat the gas around the SF region and prevents subsequent new SF.

\begin{figure}
\includegraphics[width=\columnwidth]{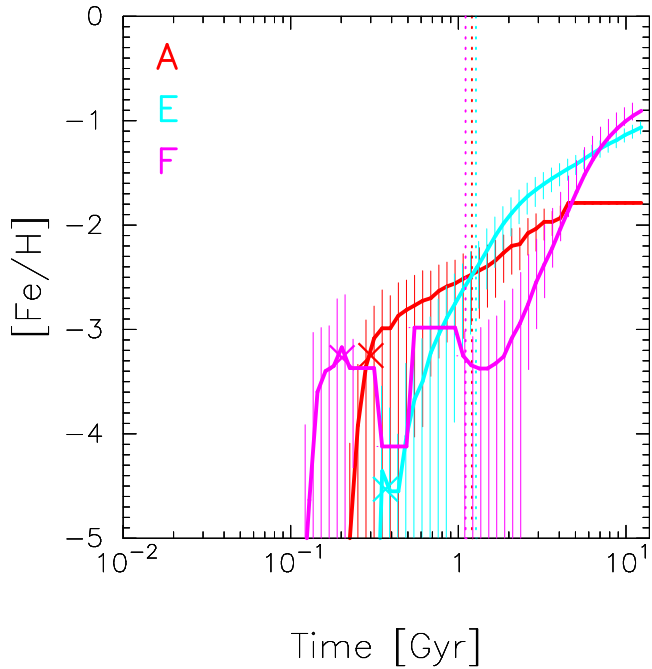}
\caption{The mean stellar [Fe/H] as a function of time (solid curves). Red, cyan, and magenta curves show models A, E, and F, respectively. Symbols and lines are the same as in Fig. \ref{AMR}. \label{AMRmass}}
\end{figure}

Fig. \ref{AMRmass} displays the mean stellar [Fe/H] as a function of time. The average metallicities at 100 Myr, [Fe/H]$_0$, in models A, E, and F are [Fe/H]$_0<-3.1$ due to suppressed SFR in these models. The value of $f_{\rm poll}$ is equal to 0.33 (model A), 0.0023 (model E) and 0.011 (model F) at 100 Myr. This result means that the spatial distribution of metallicity has not yet homogenized in all models at this phase. As shown in Fig. \ref{AMRmass}, all models show large scatters of metallicity over one dex in this phase as indicated by vertical solid lines.  Fig. \ref{AMRmass} suggests that the rate of increase in [Fe/H] is an almost identical trend irrespective of the total mass.

\begin{figure*}
\includegraphics[width=18 cm]{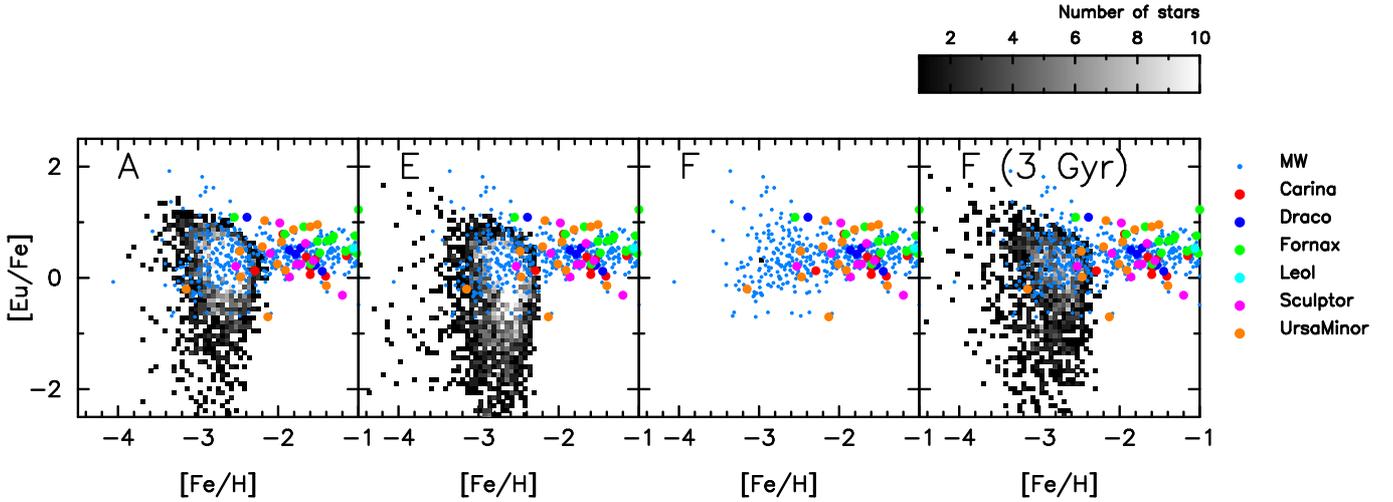}
\caption{[Eu/Fe] as a function of [Fe/H] {for stars}. From left to right, we plot models A, E, and F. We plot models in three left panels at 1 Gyr after the onset of SF. We plot model F in the rightmost panel at 3 Gyr from the start of the simulation. Symbols are the same as in Fig. \ref{EuFe1}. \label{EuFemass}}
\end{figure*}

\begin{figure}
\includegraphics[width=\columnwidth]{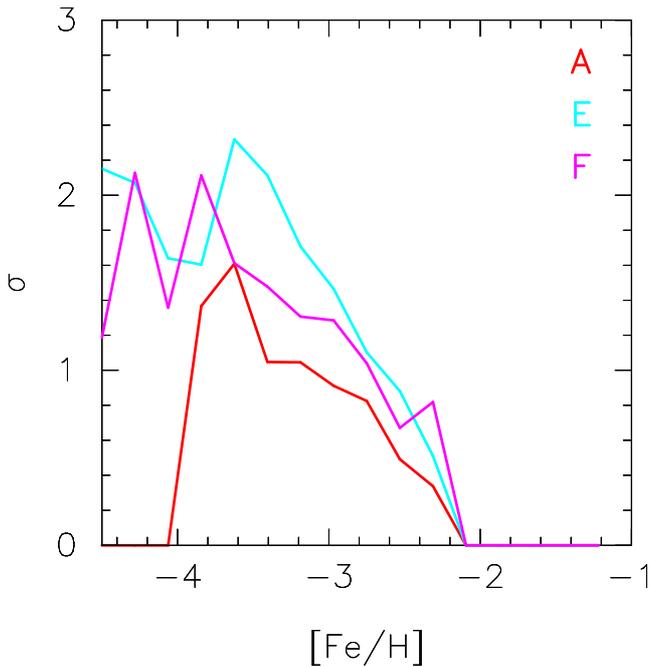}
\caption{Standard deviation of [Eu/Fe] ($\sigma$) as a function of [Fe/H] after 1 Gyr for models A, E and 3 Gyr for model F from the onset of SF. Red, cyan, and magenta curves represent models A, E, and F, respectively. \label{SigmaEuAEF}}
\end{figure}

Fig. \ref{EuFemass} shows distributions of stars in the [Eu/Fe] vs. [Fe/H] diagrams predicted in models A, E, and F.  Large dispersion in [Eu/Fe] appears at [Fe/H] $\sim~-$3 in models A and E. Fig. \ref{SigmaEuAEF} displays the standard deviation of [Eu/Fe] as a function of [Fe/H]. According to Fig. \ref{SigmaEuAEF}, models A, E, and F have $\sigma >$ 1 in [Fe/H] $\lesssim -3$. This result reflects the slow chemical evolution in the early evolutionary phase.  Model F predicts almost no EMP stars with 1 Gyr due to the suppressed SF. A sufficient number of stars which contain Eu are formed after $\sim$ 3 Gyr in model F. The distribution looks similar to those in models A and E. The major SF in model F starts after $\sim$ 2 Gyr has passed since the beginning of the simulation. The dispersion in [Eu/Fe] at 3 Gyr in model F is thus comparable to those of models A and B at 1 Gyr. These results suggest that [Eu/Fe] vs. [Fe/H] does not depend on the total mass of the system.

\subsection{Merger times of NSMs}\label{mergertime}
Here, we discuss the effect of merger times of NSMs. In the previous section, we find that the SFR in the early phase determines the metallicity at the time when the first NSM occurs. The star-to-star scatters of [Eu/Fe] decrease as the central density increases depending on the models adopted here (Fig. \ref{EuFe1}). As the central density increases, the dynamical time becomes shorter, and the star-to-star scatter of [Eu/Fe] decreases. To quantitatively examine the relation among three quantities of the distribution in [Eu/Fe] vs. [Fe/H] diagram, dynamical times, and NS merger times, we carried out simulations assuming NS merger times of 10 Myr and 500 Myr. These merger times are respectively shorter and longer than our fiducial value (100 Myr). \citet{2012ApJ...759...52D} suggest that the distribution of merger times has a power law with an index of $-1$  and the minimum time is $\sim$ 10 Myr in their binary stellar evolution models. As we mentioned in Section \ref{method}, Eu in EMP stars should come from NSMs whose merger times are shorter than the typical lifetime of SNe Ia progenitors, i.e. $\sim$ 1 Gyr. We thus adopt NS merger times of 10 and 500 Myr for models A, B, C, and D. 

\begin{figure*}
\includegraphics[width=18 cm]{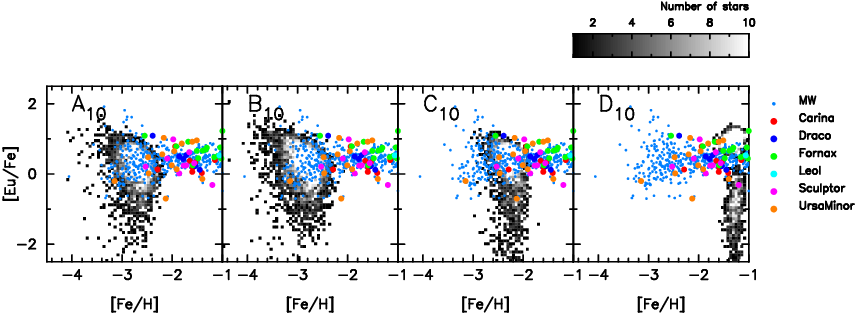}
\caption{[Eu/Fe] as a function of [Fe/H] {for stars} produced by NSMs with a merger time of 10 Myr after 1 Gyr from the onset of SF. From left to right, we plot models A$_{10}$, B$_{10}$, C$_{10}$, and D$_{10}$. Symbols are the same as Fig. \ref{EuFe1}. \label{EuFe10Myr}}
\end{figure*}

Fig. \ref{EuFe10Myr} shows [Eu/Fe] as a function of [Fe/H] for models A$_{10}$, B$_{10}$, C$_{10}$, and D$_{10}$ with NS merger times of 10 Myr. The star-to-star scatters of [Eu/Fe] in models A$_{10}$ and B$_{10}$ are not significantly different from each other as those in Fig. \ref{EuFe1}. As shown in Fig. \ref{EuFe10Myr}, a large scatter of [Eu/Fe] is seen at [Fe/H] $\lesssim -$2 in model C$_{10}$ when an NS merger time is assumed to be 10 Myr. However, it is difficult to explain Eu in low metallicity region ([Fe/H] $\lesssim -$3). In the case of model D$_{10}$, it does not explain Eu at all in [Fe/H] $<-$2. According to Fig. \ref{AMR}, the average metallicities for models C$_{10}$ and D$_{10}$ reach [Fe/H] $>-$ 2.5 in the first 10 Myr from the beginning of the simulation. It suggests that Fe increases too fast in these models to explain the dispersion in [Eu/Fe] of EMP stars.

\begin{figure*}
\includegraphics[width=18 cm]{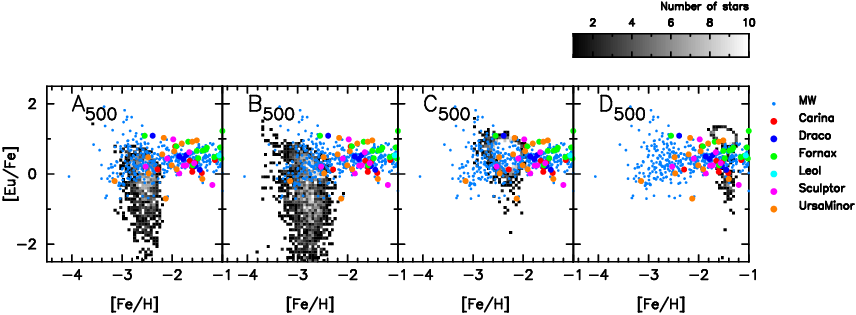}
\caption{[Eu/Fe] as a function of [Fe/H] {for stars} produced by NSMs with a merger time of 500 Myr after 1 Gyr from the onset of SF. From left to right, we plot models A$_{500}$, B$_{500}$, C$_{500}$, and D$_{500}$. Symbols are the same as Fig. \ref{EuFe1}. \label{EuFe500Myr}}
\end{figure*}

Fig. \ref{EuFe500Myr} shows [Eu/Fe] as a function of [Fe/H] for NS merger times of 500 Myr for models A$_{500}$, B$_{500}$, C$_{500}$, and D$_{500}$. The dispersions of [Fe/H] in models A$_{500}$ and B$_{500}$ are still over one dex at 500 Myr from the beginning of SF (Fig. \ref{SFR}). The large scatters of [Eu/Fe] therefore appear even if we assume an NS merger time as 500 Myr. Models C$_{500}$ and D$_{500}$ in Fig. \ref{EuFe500Myr} likewise show that [Eu/Fe] distribution is not strongly altered if we assume a merger time of 500 Myr.

From these results, we find that star-to-star scatters of $r$-process elements are not strongly affected by the merger time in the range 10 -- 500 Myr. This result suggests that even if NSMs have a distribution of merger times, the distribution of [Eu/Fe] in EMP stars does not significantly change as shown in the previous studies \citep{2015ApJ...807..115S, 2015ApJ...814...41H}.

\section{Implications to the formation of the MW halo}\label{MW}
As discussed in Section \ref{r-process}, models with low SFR ($\lesssim ~10^{-3}~M_{\odot}$ yr$^{-1}$) at the beginning can reproduce the large scatters in [Eu/Fe] of EMP stars. Therefore, if the EMP stars in the MW halo formed in accreting sub-haloes, the SFR in the sub-haloes would be suppressed such as in models A, B, E, and F. On the other hand, the average metallicity of the MW halo is [Fe/H] = $-$1.6 \citep{1991AJ....101.1865R}. Also, the constant $\alpha$-element to iron ratio in the MW halo stars ([$\alpha$/Fe] $\approx$ 0.5) indicates no contribution of SNe Ia. This observation means that the metallicity of the MW halo reaches [Fe/H] = $-$1.6 before beginning of the contribution of SNe Ia ($\sim$ 1 Gyr). According to Figs. \ref{AMR} and \ref{AMRmass}, the average metallicity in models A, B, E, and F is [Fe/H] $\lesssim~-$2.5 at 1 Gyr. This result means that the contribution of SNe Ia starts to occur at [Fe/H] $\sim~-$2 in models A, B, E, and F. This contradicts the observed [$\alpha$/Fe] abundance in the MW halo. Assembly of galaxies as in models A, B, E, and F is not possible to explain the whole metallicity range of the MW halo. It is thus required to consider other reasons to explain the stars with [Fe/H] $>~-$2 in the MW halo.

In this study, we adopt isolated dwarf galaxy models. This assumption would underestimate the rate of increase in metallicity. The main halo of the MW should have encountered mergers of sub-haloes in the context of hierarchical structure formation. This event may induce high SFRs and increase the metallicity later forming EMP stars.  The star-to-star scatters of $r$-process elements would be formed in haloes like models A, B, E, and F. Stars with [Fe/H] $>-$2 observed in the MW halo would be formed in haloes like model C. It is possible to prove whether these events really occur or not by analysing MW formation simulations with higher resolution. We will discuss this in our forthcoming paper.

\section{Summary}\label{Summary}
We performed a series of $N$-body/SPH simulations of the chemo-dynamical evolution of dwarf galaxies by varying the initial central density and the total mass of the models by one order of magnitude. We find that the distribution of $r$-process elements in EMP stars significantly depends on the dynamical evolution of galaxies reflecting different dynamical structures of galaxies (Fig. \ref{EuFe1}). The initial density, which determines the dynamical time ($t_{\rm dyn}$) of galaxies, is the most important fundamental parameter to determine the distribution of stars in [Eu/Fe] vs. [Fe/H]. The dynamical times of dwarf galaxies significantly affect the time variations of SFRs (Fig. \ref{SFR}). Models with longer dynamical times tend to form stars more slowly. We find that the early SFR is suppressed to be less than $10^{-3}~M_{\odot}$yr$^{-1}$ in galaxies with dynamical times of $\sim$ 100 Myr. If the value of $t_{\rm dyn}$ is similar, such models show a similar distribution of [Eu/Fe] vs. [Fe/H] regardless of the total mass, which is consistent with the observation of EMP stars. With a higher initial density of galaxies, in contrast, [Eu/Fe] distribution shifts to a higher metallicity and the star-to-star scatters diminish. {We confirm that the scatters of [Eu/Fe] mainly come from the inhomogeneity of the metals in the ISM whereas the scatters of $\alpha$-elements are predominantly due to the difference in the yield of each SNe.}

These results suggest that the observed star-to-star scatters of [Eu/Fe] in EMP stars in the MW halo can be explained by NSMs if the MW halo formed via accretion of dwarf galaxies with the initial central density of $<10^8~M_{\odot}$kpc$^{-3}$. Our results also imply that [Eu/Fe] distribution in individual galaxies reflects the early SFR and dynamical times of galaxies. 

We also find that NS merger times between 10 Myr and 500 Myr do not strongly affect the final results. Low-density models such as models A and B can predict star-to-star scatters in [Eu/Fe] even if we assume an NS merger time of 500 Myr (Fig. \ref{EuFe500Myr}). To explain the final abundance, we do not need to adopt unlikely short merger times if the SFR is suppressed to be less than $10^{-3}~M_{\odot}$yr$^{-1}$.

We expect that future precise observations of EMP stars in dwarf galaxies will prove the early evolutionary history of the LG galaxies. If the star-to-star scatters of [Eu/Fe] would appear in galaxies at similar metallicities with different stellar mass, the SFR of galaxies should be lower than $10^{-3}~M_{\odot}$yr$^{-1}$ regardless of their total mass. On the other hand, if the smaller scatters would appear at higher metallicities in galaxies with higher stellar mass, this could be a signal that the early SFRs for heavier galaxies is higher due to higher initial central density. This study suggests that the future high-dispersion spectroscopic observations of the abundance of $r$-process elements will be able to constrain the early evolutionary histories of LG galaxies.

\section*{Acknowledgements}
We are grateful to an anonymous reviewer who gives us insightful comments to improve the paper greatly. We thank Takuma Suda for providing us with the latest version of SAGA database including the dataset of the LG dwarf galaxies and fruitful discussion. This work was supported by JSPS KAKENHI Grant Numbers 15J00548, 26707007, 26800108, 26105517, 24340060, 15H03665, MEXT SPIRE and JICFuS. This study was also supported by JSPS and CNRS under the Japan-France Research Cooperative Program. Numerical computations and analysis were in part carried out on Cray XC30 and computers at Centre for Computational Astrophysics, National Astronomical Observatory of Japan. This research has made use of NASA's Astrophysics Data System.\newpage

\bibliographystyle{mnras}
\bibliography{sampleNotes}

\bsp	
\label{lastpage}
\end{document}